\documentclass[a4paper,10pt]{article}
\usepackage{amsmath, amssymb, graphics,graphicx}
\usepackage{amscd}
\usepackage[dvips]{hyperref}
\usepackage{textcomp}
\def\complex{{\mathbb  C}}
\def\reals{{\mathbb  R}}

\def\vvec{\underline{v}}
\def\uvec{\underline{u}}
\def\rvec{\underline{r}}
\def\omvec{\underline{\omega}}
\def\del{\underline{\nabla}\,}
\def\delsq{\nabla^2}
\def\delf{\nabla^4}
\def\Psivec{\underline{\Psi}}
\def\exvec{\underline{e}_x}
\def\eyvec{\underline{e}_y}

\begin{document}
\title{Complex Variable Methods for 3D Applied Mathematics: \\ 3D Twistors and the biharmonic equation\thanks{Working Paper V1.2, May 2010}}
\author{William T. Shaw\thanks{Department of Mathematics, King's College, The Strand, London WC2R 2LS}}

\maketitle
\begin{abstract}
In applied mathematics generally and fluid dynamics in particular, the role of complex variable methods is normally confined to two-dimensional motion and the association of points with complex numbers via the assignment $w = x+i y$. In this framework 2D potential flow can be treated through the use of holomorphic functions and biharmonic flow through a simple, but superficially non-holomorphic extension. This paper explains how to elevate the use of complex methods to three dimensions, using Penrose's theory of twistors as adapted to intrinsically 3D and non-relativistic problems by Hitchin. We first summarize the equations of 3D steady viscous fluid flow in their basic geometric form. We then explain the theory of twistors for 3D, resulting in complex holomorphic representations of solutions to harmonic and biharmonic problems. It is shown how this intrinsically holomorphic 3D approach reduces naturally to the well-known 2D situations when there is translational or rotational symmetry, and an example is given. We also show how the case of small but finite Reynolds number can be integrated by complex variable techniques in two dimensions, albeit under strong assumptions.  
\end{abstract}

\noindent
Key Words: Fluid dynamics, Twistor theory, Three dimensions, Stokes flow, Complex variables, Stream function, Biharmonic equation, Low Reynolds number

\section{Introduction}
The use of complex variable techniques in applied mathematics, and especially fluid dynamics, is dominated by two-dimensional applications through the prescription
\begin{equation}
w = x + i y \ .
\end{equation}
On the other hand, in the context of relativistic physics in four or more dimensions, the use of twistor methods due to R. Penrose and co-workers is becoming an ever more present tool in the hands of theoretical physicists. Many of the key earlier developments are summarized by Penrose \& Rindler (1984). A collection of more recent developments is available (Mason \& Hughston, 1990, Mason {\it et al.}, 1990) and some state of the art applications to superstring theory are given in a very recent thread of work initiated by Witten (2003) (see also the thread on the web site \url{http://arxiv.org/cits/hep-th/0312171)}.

In the important intermediate case of non-relativistic problems in three spatial dimensions, rather than working with a copy of $\complex$ for the two-dimensional case, or with $\complex P^3$ (the twistor space for Minkowski space-time) we work with the twistor space for three dimensions: $T\complex P^1$ {\textendash} the tangent space for the Riemann sphere. This case has not received quite such wide attention. A notable exception is the paper by N. Hitchin (1982), where a summary of the basic theory appeared as a preparation for an application to monopoles. In a notable Appendix, Hitchin (1982) also explained the use of $T\complex P^1$ to explain the geometrical underpinnings of Weierstrass's holomorphic solution of Plateau's problem (minimal surfaces, or soap bubbles) (Weierstrass, 1866). Hitchin lectured in Oxford and Stony Brook on the methods in the 1980s, but those talks have not been published. A summary of the some of basic ideas accessible to non-specialists is now available (Shaw, 2006), based on developing the three-dimensional theory via the relativistic method, and including applications to Laplace's equation in three dimensions. These methods give the complex geometrical underpinnings of contour integral formulae obtained by more {\it ad hoc} methods and known for some time - see for example Whittaker's work (Whittaker, 1903). The applications of this approach to various types of problem with axis-symmetry has been given by Mason (1990) and Ward (1981).

The focus of much of the published work has been on time-independent problems within the general context of theoretical relativistic physics. In this paper the idea is to present such methods as being a routinely useful tool in traditional applied mathematics. To this end, an example of the application of twistor theory to viscous fluid flow is presented. In particular, the solution of various biharmonic problems will be presented using contour integral techniques. The ultimate goal of this work is a better understanding of the Navier-Stokes equations through the geometry of {\it holomorphic} complex variable techniques. At first sight, even our most basic goal might seem to be an unreasonable proposal. For example, the biharmonic equation in two dimensions, with the $w = x + i y$ prescription, amounts to
\begin{equation}
\partial_w^2 \partial_{\bar{w}}^2 \Psi = 0
\end{equation}
with the general real solution
\begin{equation}
\Psi = \Re\bigl\{\bar{w} f(w) + g(w)\bigr\}
\end{equation}
where $f$ and $g$ are both locally holomorphic. This is generally regarded as going outside the holomorphic context as it involves $\bar{w}$ in an essential way. We shall show that equation (1.3) is in fact the two-dimensional projection of an {\it essentially holomorphic} three-dimensional result. 

One goal of this work is a better understanding of the Navier-Stokes equations through the geometry of complex variable techniques. There is already progress in this area. A characterization of the full steady problem in 2D has been given by Legendre (1949) and Ranger (1991, 1994). Work on the K\"{a}hler Geometry associated with the Navier-Stokes problem has been given by Roulstone {\it et al} (2005), as well as related characterizations Roulstone {\it et al} 2009. It would give considerable insight to draw together theses various threads. We cannot expect to be able to analyze the full Navier-Stokes system in three dimensions using twistor methods if we cannot first treat the biharmonic limit, and this may well give insight into the full problem {\textendash} this current paper suggests it is at least straightforward to treat the biharmonic case. 

There are also several ways of attacking the problem based on different choices of physical equation that we wish to transform into a holomorphic geometrical description. For example, given that the momentum equation is fundamental, one could work with the equation directly, or its curl, or its divergence. Here we work with the curl in the guise of the vorticity equation. Roulstone {\it et al} (2005) work with the divergence. Similarly, here we work with a vector potential for the velocity. One might also consider working with the Clebsch potential representation, as discussed, for example, by Milne-Thomson (see Section 21.021 of Milne-Thomson, 1996). Generalizations of this approach are much favoured for the treatment of fluids by action principles (see, e.g. Morrison, 1998). We will take the vector potential approach rather than the  ``Clebsch'' route here for three reasons. First, in the latter approach the vorticity is already a non-linear function of the basic variables. Second, the relationship between action principles and twistor methods is an uneasy one, since the power of twistor methods lies mainly (so far) in an understanding of how the differential equations governing a system may be solved exactly by the introduction of some holomorphic structure. Dynamical configurations that are ``off-shell'' (in physics terminology), i.e. are not solutions of the equations of motion in space or space-time, do not so far appear to have a straightforward twistor description, though the recent work by Witten (2003) makes use of action functionals in twistor space to describe particle interactions. Finally, in taking the vector potential approach, we are able to make contact, for systems with certain symmetry, with the simple stream function approach. 

A key outcome of the development given here is that one can show show that there is a simple holomorphic framework for three dimensions, within which we can treat the vector biharmonic equation in holomorphic terms. We also show how to recover not only Eq. (1.3) for systems with translational symmetry, but also a contour integral representation of the Stokes stream function for the axis-symmetric case. These two symmetric systems emerge from a {\it common 3D holomorphic framework}. 

Although we are working  with a biharmonic system motivated by fluid dynamics, there are of course applications to elasticity with regard to both the biharmonic equation in general (see Note 93 of Love, (1927) and Howell {\it et al.}, (2008) for a modern treatment) and specifically complex variable methods (Muskhelishvili, 1949,  England, 1971). The methods discussed in this paper can equally be applied to the three-dimensional Lam\'{e} or Navier equation for steady configurations of incompressible elastic media under a potential body force.  

The plan of the work is as follows. In Section 2 a brief summary of viscous fluid dynamics is given. In Section 3 a self-contained introduction to twistors for three dimensions is given together with the basic ideas of how to solve the scalar Laplace and biharmonic equations. In Section 4 this is extended to the vector biharmonic case and the relationship to planar and axis-symmetric stream functions is elucidated.  In Section 5 we take a first look at the non-linear problem arising from the case of small but finite Reynolds number in two dimensions. It is shown how this may be integrated (almost explicitly) using entirely holomorphic methods, under strong assumptions on the nature of the flow.

\section{Steady viscous incompressible flow}
A large class of fluids can be characterized by their density, $\rho$, a scalar field not presumed to be constant, and their dynamic viscosity $\mu$. The flow is characterized by a velocity vector field $\underline{v}$, and an associated scalar pressure field $p$. Conservation of mass is expressed by the continuity equation
\begin{equation}
\frac{\partial \rho}{\partial t} + \del.(\rho \vvec) = 0 
\end{equation}
and the conservation of momentum is expressed by the Navier-Stokes equations\footnote{In Sections 2 and 3 of this article, $\delsq$ acting on {\it vectors} should be understood as the ordinary Laplacian acting on Cartesian components. The effects of a non-constant basis are dealt with in Section 4.}
\begin{equation}
\rho(\frac{\partial \vvec}{\partial t} + \vvec.\del \vvec) = -\del p + \mu \delsq \vvec
\end{equation}
If the fluid is incompressible in the sense that $\rho$ is a constant in both time and space, we have the condition:
\begin{equation}
\del.\vvec = 0
\end{equation}
To analyze matters further, we introduce the vorticity vector
\begin{equation}
\omvec = \del \times \vvec
\end{equation}
In the following discussion we demand incompressibility but allow for non-zero vorticity. Using simple identities from vector calculus the Navier-Stokes equations may then be recast in the form
\begin{equation}
\rho(\frac{\partial \vvec}{\partial t} - \vvec \times \omvec) + \del (p + \frac{1}{2} \rho \vvec^2
)= -\mu \del \times \omega .
\end{equation}
Taking the curl of this, we arrive at the vorticity equation
\begin{equation}
\frac{\partial \omvec}{\partial t} + \vvec.\del \omvec - \omvec.\del \vvec = \nu \delsq \omvec
\end{equation}
where the kinematic viscosity $\nu = \mu/\rho$.
\subsection{The `stream vector potential'}
Since the velocity field is divergence-free, we may introduce a vector potential $\Psivec$ such that
\begin{equation}
\vvec = \del \times \Psivec
\end{equation}
and furthermore we may choose it so that it is divergence free:
\begin{equation}
\del.\Psivec = 0
\end{equation}
In theoretical physics, notably electromagnetic theory, this is known as setting a gauge condition.
The tradition in fluid dynamics is to mainly use the vector potential only when it can be reduced to a single function using some type of symmetry. The resulting object is a stream function. For example, planar 2D flow is obtained by setting (and note that this automatically satisfies satisfies the gauge condition)
\begin{equation}
\Psivec = - \psi(x, y) \underline{e}_z
\end{equation}
We will work with the full vector form. First of all we note that under the assumption that $\Psi$ satisfies $\del.\Psivec = 0$ 
\begin{equation}
\omega = -\delsq \Psivec
\end{equation}
and the vorticity equation becomes, denoting $\frac{\partial\ }{\partial t}$ by $\dot{\ }$:
\begin{equation}
\nabla^4 \Psivec=\frac{1}{\nu}\biggl\{((\delsq\Psivec).\del) \del \times \Psivec-((\del \times \Psivec).\del)(\delsq \Psivec) + \delsq \dot{\Psivec}\biggr\}
\end{equation}
or indeed as
\begin{equation}
\nabla^4 \Psivec = \frac{1}{\nu}\biggl\{\del \times ((\del \times \Psivec)\times \delsq \Psivec)+\delsq \dot{\Psivec}\biggr\}
\end{equation}
This latter representation of the Navier-Stokes equations is well-known in the 2D planar case (see for example Ockendon {\it et al.} 2003) where it reduces, in the time-independent case, to the equation
\begin{equation}
\nabla^4 \psi = \frac{1}{\nu} \frac{\partial(\psi, \delsq \psi)}{\partial(x, y)}
\end{equation}
\subsection{The biharmonic limit}
For problems where it is possible to identify a natural length scale $L$ and a natural speed $U$, it is normal practice to perform a non-dimensionalization of the variables and introduce the Reynolds number $R = UL/\nu$. Then the steady-state vorticity equation becomes (after rescaling the independent variables suitably):
\begin{equation}
\nabla^4 \Psivec =R \biggl\{\del \times ((\del \times \Psivec)\times \delsq \Psivec)\biggr\}
\end{equation}
The old historical approach to the limiting case when $R \rightarrow 0$ is to take the view that the non-linearities may be ignored (provided the non-linear term is well behaved) and the time-independent Navier-Stokes equations reduce to 
\begin{equation}
\nabla^4 \Psivec = 0
\end{equation}
which is the biharmonic limit, also known as Stokes flow. We shall focus on the solution of this equation by complex variable methods. It is now  well known (see for example, Chapter 8 of Van Dyke (1964)) that the neglect of the non-linear terms can lead to inconsistencies, as is evidenced by the lack of any solution for asymptotically uniform two-dimensional flow past a cylinder. However, in attempting to construct a twistor description of fluid flow we {\it must} be able to at least solve the biharmonic equation. It is to this that we now turn. 
\section{Twistor solutions of the Laplace and biharmonic equation}
It is very well known that the Laplace equation can be solved in terms of holomorphic functions in two dimensions. Among devotees of twistor methods, it is also well known that this can be carried out in three dimensions. Our purpose is to extend this idea to the biharmonic case. 
\subsection{Twistor space}
The twistor solution of the Laplace equation in three dimensions may be discussed at several levels. In presenting this material the simplest and hopefully most comprehensible route shall be  taken, but it is necessary to be somewhat formal in order to state results correctly. This presentation follows that given by Hitchin (1982)\footnote{This presentation also draws on Hitchin's unpublished lectures, particularly with regard to the solution of Laplace's equation and understanding the 2D limit}. The twistor space associated with $\reals^3$ is first, as a real space, the set of oriented straight lines in $\reals^3$. Relative to some origin $O$, let $\underline{r}$ denote the position vector of the point on a given line nearest to $O$. Then $\underline{r}$ is orthogonal to the direction of the line, which we denote by $\underline{u}$ with $\underline{u}.\underline{u} = 1$. So the set of oriented straight lines is the set
\begin{equation}
TS^2 = \bigg\{ (\underline{r}, \underline{u}) \in \reals^3 \times S^2 \ | \ \underline{r}.\underline{u} = 0 \biggr\}
\end{equation}
where our notation emphasizes that this is the tangent bundle for the unit sphere.
This set is also naturally the tangent bundle to a complex manifold, where we regard $S^2$ as the Riemann sphere $\complex P^1$. This complex tangent bundle, $T\complex P^1$ is the twistor space of interest. 

The next question is how to define a point in ordinary space in terms of some structure on $T\complex P^1$? A point may be regarded as the intersection of all straight lines through it. This means that a point is necessarily some vector field in $T\complex P^1$ that is defined {\it globally}. To see the implications of this we introduce two open sets that cover $\complex P^1$. We can take coordinates for the sphere as $\zeta$ on one patch (covering everything except infinity), and $\tilde{\zeta} = 1/\zeta$ on another patch, covering everything except $\zeta = 0$.  Over each of these respective patches we can define coordinates for the tangent bundle as $(\eta, \zeta)$ and $(\tilde{\eta}, \tilde{\zeta})$, where the relevant vector fields are, respectively
\begin{equation}
\eta \frac{\partial\ }{\partial \zeta}\ \ , \ \ \ \ \tilde{\eta} \frac{\partial\ }{\partial \tilde{\zeta}}
\end{equation}
Consider now a holomorphic vector field. On the $\zeta$ patch it can be written as 
\begin{equation}
f_0(\zeta)\frac{\partial\ }{\partial \zeta}
\end{equation}
for some $f_0$, and on the $\tilde{\zeta}$ patch, it can be written as
\begin{equation}
f_1(\tilde{\zeta})\frac{\partial\ }{\partial \tilde{\zeta}}
\end{equation}
for some $f_1$. On the intersection of the two patches equality of the two representations gives us
\begin{equation}
f_1(\zeta^{-1})(-\zeta^2) \frac{\partial\ }{\partial \zeta}
= f_0(\zeta)\frac{\partial\ }{\partial \zeta}
\end{equation}
If we make a Taylor series expansion of both functions, $f_i(\zeta) = \sum_{n=0}^{\infty} a_n^i \zeta^n$, we deduce that the coefficients $a_n^i$ vanish if $n>2$. That is, the global vector fields must be of the form, for example on the $\eta$ patch:
\begin{equation}
\eta(\zeta) = a + b \zeta + c \zeta^2
\end{equation}
so that such quadratics are the only holomorphic vector fields, and these correspond to points of $\complex^3$, parametrized in some way by $(a,b,c)$.

Further analysis of this system allows the identification of real points in $\reals^3$, and the construction of a natural metric. The points are real if and only if 
\begin{equation}
c = -\overline{a} \ \ \ {\rm AND}\ \ \ b = \overline{b}
\end{equation}
The induced metric is proportional to the discriminant of the quadratic, and we shall normalize matters such that
\begin{equation}
ds^2 = dx^2 + dy^2 + dz^2 = \frac{1}{4} db^2 - da dc
\end{equation}
The metric for quadratics corresponding to real points is therefore:
\begin{equation}
ds^2 = dx^2 + dy^2 + dz^2 = \frac{1}{4} db^2 + da d\overline{a}
\end{equation}
If we pick our coordinate system such that the real part of $a$ is $x$, we see that we can take the imaginary part of $a$ to be $\pm y$ and set $b = \pm 2 z$. The convention is to set:
\begin{equation}
\eta_{\underline{r}}(\zeta) = (x+ iy) + 2 z \zeta - (x-i y) \zeta^2
\end{equation}
Note that here and elsewhere we will carefully avoid {\it ever} writing $z = x+iy$! Note that although this particular Cartesian representation does not make manifest the action of rotations on $\reals^3$ it will turn out to be vary useful to deal with the cases of translational and axial symmetry. This will be discussed in Sub-sections 2(d) and 2(e). 

\subsection{Solving the scalar Laplace equation}
We consider a function $f(\eta, \zeta)$ defined on twistor space. This can then be thought of as restricted to the special global sections of twistor space represented by $\eta_{\underline{r}}(\zeta)$, and the $\zeta$-dependence integrated out by integration over a contour $C$. We set:
\begin{equation}
\phi(\underline{r}) = \int_C f(\eta_{\underline{r}}(\zeta), \zeta) d\zeta
\end{equation}
Then $\phi$ satisfies the scalar Laplace's equation. To see this observe that
\begin{equation}
\frac{\partial^k f(\eta_{\underline{r}}(\zeta), \zeta)}{\partial x^k} = (1-\zeta^2)^k \frac{\partial^k f}{\partial \eta^k}|_{\eta = \eta_{\underline{r}}}
\end{equation}
\begin{equation}
\frac{\partial^k f(\eta_{\underline{r}}(\zeta), \zeta)}{\partial y^k} = i^k (1+\zeta^2)^k \frac{\partial^k f}{\partial \eta^k}|_{\eta = \eta_{\underline{r}}}
\end{equation}
\begin{equation}
\frac{\partial^k f(\eta_{\underline{r}}(\zeta), \zeta)}{\partial z^k} = (2 \zeta)^k \frac{\partial^k f}{\partial \eta^k}|_{\eta = \eta_{\underline{r}}}
\end{equation}
and that adding these three expressions with $k=2$ gives zero identically for any choice of $f$. Note that many different choices of $f$ will give rise to the same $\phi$. Such choices differ by the additions of functions that are holomorphic inside or outside of C, so that one must pursue a cohomological approach in order to state a formal isomorphism between structures on twistor space and solutions of the Laplace equation. 
\subsection{Solving the scalar biharmonic equation}
The extension of the results for the scalar Laplace equation to the scalar biharmonic case may be pursued on several levels. Our argument does not require any advanced knowledge of twistor theory, though it would be interesting to characterize this analysis in terms of the appropriate cohomology structures. We shall pose the following question. How do we modify the integrand  $f(\eta_{\underline{r}}(\zeta), \zeta)$, say to some holomorphic function $g$, to arrange that $\delf g = 0$ but $\delsq g \neq 0$? 

Let's consider trying to build $g$ from $f$ by multiplying by some prefactor $h(\underline{r}, \zeta)$, so that
\begin{equation}
g = h(\underline{r}, \zeta) f(\eta_{\underline{r}}(\zeta), \zeta)
\end{equation}
Now 
\begin{equation}
\delsq g = \delsq h f = f \delsq h + h \delsq f + 2 \del h.\del f = f \delsq h + 2 \del h.\del f
\end{equation}
where the last simplification arises as $f$ satisfies the Laplace equation. If we furthermore choose $h$ to be linear in $\underline{r}$ matters simplify further and we have
\begin{equation}
\delsq g = 2 \del h.\del f
\end{equation}
Let us set, w.l.o.g. (other than excluding $h$ being constant, which gives us harmonic solutions already understood),  $h = \uvec(\zeta).\rvec$, so that $\del h = \uvec(\zeta)$. We also note that
\begin{equation}
\del f = \frac{\partial f}{\partial \eta}\del \eta = \frac{\partial f}{\partial \eta} (1-\zeta^2, i(1+\zeta^2), 2 \zeta).
\end{equation}
Putting this all together, we arrive at
\begin{equation}
\delsq g = \delsq h f = 2 \uvec(\zeta).(1-\zeta^2, i(1+\zeta^2), 2 \zeta) \frac{\partial f}{\partial \eta} = 2 \eta_{\uvec(\zeta)}(\zeta)\frac{\partial f}{\partial \eta}
\end{equation}
We can now see that $\delf g=0$ identically, while $\delsq q$ does not vanish unless $\eta_{\uvec(\zeta)}(\zeta) \equiv 0$. To see what is happening, we can now make matters more explicit. We let $\uvec(\zeta) = (u_1(\zeta),u_2(\zeta),u_3(\zeta))$, then
\begin{equation}
\uvec.\rvec = u_1(\zeta) x + u_3(\zeta) y + u_3(\zeta) z
\end{equation}
and
\begin{equation}
\eta_{\uvec(\zeta)}(\zeta) = (u_1(\zeta) + i u_2(\zeta)) + 2 u_3(\zeta) \zeta - (u_1(\zeta) - i u_2(\zeta))\zeta^2
\end{equation}
In terms of these variables the proposed integral representation for solutions of the 3D {\it scalar} biharmonic equation is just
\begin{equation}
\Psi = \int_C \! d \zeta \biggl[
x u_1(\zeta) +y u_2(\zeta)) +  z u_3(\zeta)
\biggr]  f(\eta_{\underline{r}}(\zeta), \zeta)
\end{equation}
or indeed as (this  will prove  useful presently  when looking at axis-symmetry):
with $w = x + i y$:
\begin{equation}
\Psi = \frac{1}{2}\int_C \! d \zeta \biggl[w g_-(\zeta) + \bar{w}g_+(\zeta) + 2 z u_3(\zeta)\biggr]  f(\eta_{\underline{r}}(\zeta), \zeta)
\end{equation}
where $g_{\pm}(\zeta) = u_1(\zeta)\pm i u_2(\zeta))$. Note that the introduction of $\bar{w}$ is only a parametrization convenience {\textendash} it has nothing to do with the underlying holomorphic structure of twistor space. 
\subsection{The scalar biharmonic problem in 2D}
Suppose we want no $z$-dependence. We set $u_3 = 0$ and $w = x + i y$, so that
\begin{equation}
\Psi = \int_C \! d \zeta \biggl[w g_-(\zeta) + \bar{w}g_+(\zeta)\biggr]  f(\eta_{\underline{r}}(\zeta), \zeta)
\end{equation}
We can write this in the equivalent form
\begin{equation}
\Psi = w\! \int_C \! d \zeta f_1(\eta_{\underline{r}}(\zeta), \zeta) + \bar{w}\! \int_C \! d \zeta
f_2(\eta_{\underline{r}}(\zeta), \zeta)
\end{equation}
Now consider the second term. This is $\bar{w} \phi(x,y,z)$, where
$\phi$ is a solution of Laplace's equation and is just
\begin{equation}
\phi(x,y,z) = \int_C \! d \zeta f_2(\eta_{\underline{r}}(\zeta), \zeta)
\end{equation}
We want this not to depend on $z$ either. But this looks awkward given that 
$\eta_{\underline{r}}(\zeta) = (x+ iy) + 2 z \zeta - (x-i y) \zeta^2$. It is not so awkward as it looks. What we want is translation invariance for $\phi$. Note that
\begin{equation}
\phi(x,y,z+h/2) = \int_C \! d \zeta f_2(\eta_{\underline{r}}(\zeta)+ h \zeta, \zeta)
\end{equation}
The equation we need is
\begin{equation}\phi(x,y,z+h/2) = \phi(x,y,z)\end{equation}
This does {\it not} require that
\begin{equation}f_2(\eta_{\underline{r}}(\zeta)+ h \zeta, \zeta) = f_2(\eta_{\underline{r}}(\zeta), \zeta) \end{equation}
Instead we need
\begin{equation}f_2(\eta_{\underline{r}}(\zeta)+ h \zeta, \zeta) = f_2(\eta_{\underline{r}}(\zeta), \zeta)+g_0(\eta, \zeta, h)-g_1(\eta, \zeta, h) \end{equation}
where $g_0$ is holomorphic on and inside C and $g_1$ is likewise outside. Cauchy's theorem then gives us the desired result. In order to give a clear calculation. Let's take $C$ to be unit circle, or to be deformable to the unit circle. Now differentiate w.r.t $h$ then set $h=0$. We obtain, for some $G_i$, 
\begin{equation}
\zeta \frac{\partial f_2}{\partial \eta} = G_0(\eta, \zeta) - G_1(\eta, \zeta)
\end{equation}
We integrate this w.r.t. $\eta$ and divide by $\zeta$. We obtain, for some $H_i$, 
\begin{equation}
f_2 = \frac{H_0(\eta, \zeta)}{\zeta} - \frac{H_1(\eta, \zeta)}{\zeta}
\end{equation}
and recall that $H_0$ must be holomorphic inside C and $H_1$ holomorphic outside. 
Now we evaluate the integral of Eq. (3.26) using calculus of residues. The first term in Eq. (3.32) is easy, and we get
\begin{equation}
2 \pi i H_0 (\eta_{\underline{r}}(0),0) = K(w)
\end{equation}
for some function $K(w)$, giving a contribution to $\phi$ of $\bar{w} K(w)$. When we calculate the contribution of the second term of Eq. (3.32) to the integral of Eq. (3.26), we make the transformation $\zeta \rightarrow \tilde{\zeta}$ and obtain an integrand that is a function of $\tilde{\eta} = (x-i y) - 2 z \tilde{\zeta} - (x+iy)\tilde{\zeta}^2 $. Taking the residue at $\tilde{\zeta} = 0$ gives a function of $\bar{w} = x-iy$, also to be multiplied by $\bar{w}$.

The other two terms in Eq. (3.25) may be treated similarly. We end up with four terms contributing to $\Psi$: 
\begin{equation}\label{biharsoltwo}
\Psi = \bar{w} K_2(w) + \bar{w} \tilde{K}_2(\bar{w}) + w K_1(\bar{w}) + w \tilde{K}_1(w)
\end{equation}
When $\Psi$ is real we must have Eq. (1.3).  So the fully holomorphic picture in three dimensions projects, via the calculus of residues, to a two-dimensional picture and generates the familiar yet superficially non-holomorphic two-dimensional representation of solutions to biharmonic (and Laplace) equations. In three dimensions our functions are contour integrals. 

\subsection{The axis-symmetric scalar problem}
This problem is one of considerable interest, though it is important to realize that this is a different problem from the case of vectorial axis-symmetric flow {\textendash} that issue and the link to the Stokes stream function will be discussed later. We go back to the representation
\begin{equation}
\Psi = \frac{1}{2}\int_C \! d \zeta \biggl[w g_-(\zeta) + \bar{w}g_+(\zeta) + 2 z u_3(\zeta)\biggr]  f(\eta_{\underline{r}}(\zeta), \zeta)
\end{equation}
with $w = x + i y$. We can regard this as three pieces, where we discard the factors of a half:
\begin{equation}
\Psi_- = w \int_C \! d \zeta g_-(\zeta) f(\eta_{\underline{r}}(\zeta), \zeta) = w \psi_-
\end{equation}
\begin{equation}
\Psi_+ = \bar{w}\int_C \! d \zeta g_+(\zeta) f(\eta_{\underline{r}}(\zeta), \zeta) = \bar{w} \psi_+
\end{equation}
\begin{equation}
\Psi_3 = z \int_C \! d \zeta u_3(\zeta)  f(\eta_{\underline{r}}(\zeta), \zeta) = z \psi_0
\end{equation}
In order to develop axis-symmetric solutions, we need to understand the action of the group of rotations about the $z$-axis. We need to bear in mind the formula
\begin{equation}
\eta = w + 2 z \zeta - \bar{w} \zeta^2
\end{equation} 
with $w = x + i y$. Under a rotation about the $z$-axis, $z \rightarrow z$ and $w \rightarrow \exp(i\phi) w$. This is compatible with the action $(\eta, \zeta) \rightarrow \exp(i\phi) (\eta, \zeta) $. In seeking axis-symmetric solutions for $\psi_{\pm,0}$ we need to arrange that
\begin{equation}
d\zeta g_- f \rightarrow \exp(-i\phi) d\zeta g_- f 
\end{equation}
\begin{equation}
d\zeta g_+ f \rightarrow \exp(i\phi) d\zeta g_+ f 
\end{equation}
\begin{equation}
d\zeta u_3 f \rightarrow  d\zeta u_3 f 
\end{equation}
To treat all of these situations together, we consider the case where $d\zeta h(\eta, \zeta) \rightarrow \exp(im\phi)d\zeta h(\eta, \zeta)$. To this end we consider the contour of integration to be the unit circle and consider a basic set
\begin{equation}
\psi_{n,m} = \frac{1}{2 \pi i} \int d \zeta \frac{\eta^n}{\zeta^{n+1-m}}
\end{equation}
where for $\psi_0$, $m=0$, and for $\phi_{\pm}$, $m=\pm 1$. So our task now is to calculate
\begin{equation}
\psi_{n,m} = \frac{1}{2 \pi i} \int d \zeta \frac{1}{\zeta^{n+1-m}}(w+2z\zeta-\bar{w} \zeta^2)^n
\end{equation}
By multiplying these by the relevant factors of $w, \bar{w}, z$ for $m = -1, 1, 0$ we get an interesting set of axis-symmetric biharmonic functions. The functions $\psi_{n,m}$.  themselves are now contour integral solutions of Laplace's equation. This is of course of interest in itself. 

Note that a clue to what is going to emerge is that if $n = 1,2,3\dots$, it is easily seen that the integral for $\psi_{n,m}$ will vanish if there is a zero coefficient of $\zeta^{-1}$, and this will happen if $m>n$ or $m<-n$. So with $n>0$ we are only interested in $|m| \leq n$ (this is a hint to what special functions might emerge). 

To evaluate this set we let $y=0$ since $\psi_{n,m}(r,\theta,\phi) = e^{im\phi}\psi(r, \theta, 0)$. Then, we have, in spherical polar coordinates, 
\begin{equation}
\psi_{n,m} = \frac{1}{2 \pi i} \int d \zeta \zeta^{m-1}(2\cos(\theta)+(\frac{1}{\zeta}-\zeta)\sin(\theta))^n
\end{equation}
Parametrizing the integral as $\zeta = e^{it}$, we obtain
\begin{equation}
\psi_{n,m} = \frac{(2 r)^n}{2 \pi} \int dt  e^{imt}(\cos(\theta) - \sin(\theta)\sin(t))^n
\end{equation}
On consulting Gradstheyn and Rhyzik (1980) equation 8.711.2, and performing some manipulations, we see that, discarding normalizations, if $n\neq -1$.
\begin{equation}
\psi_{n,m} \propto
\begin{cases}
r^n P_n^m(\cos(\theta)) & n=0,1,2,\dots,\\
\frac{1}{r^{k+1}} P_k^m(\cos(\theta))& n=-k-1,k=1,2,3,\dots
\end{cases}
\end{equation}
When $n=-1$ matters are quite subtle as the integral branches depending on the sign of $z$! A full treatment of this is rather beyond the scope of this paper but we note that in this case,
\begin{equation}
\psi_{-1,m} = \frac{1}{2\pi i} \int d\zeta  \frac{\zeta^m}{w + 2 z \zeta - \bar{w} \zeta^2}
\end{equation}
The quadratic in the denominator has two roots $\zeta_{\pm}$ given by
\begin{equation}
\zeta_{\pm} = \frac{-z \pm r}{-\bar{w}}\ ,\ \ r = \sqrt{x^2 + y^2 + z^2}\ ,\ \ \bar{w} = x - i y \ .
\end{equation}
These roots are located, using standard spherical polar coordinates, at
\begin{equation}
\zeta_+ = -e^{i\phi}\tan(\frac{\theta}{2})\ ,\ \ \zeta_- = e^{i\phi}\cot(\frac{\theta}{2})
\end{equation} 
and we can write
\begin{equation}
\psi_{-1,m} = \frac{-1}{2\pi i \bar{w}} \int d\zeta  \frac{\zeta^m}{\bigl((\zeta - \zeta_+)(\zeta - \zeta_-)\bigr)}
\end{equation}
The details of the global evaluation of this are lengthy. We note here that when $z>0$, $|\zeta_+|<1$ and when $m \geq 0$, the single residue inside the unit circle gives a value of  
\begin{equation}
\psi_{-1,m} = \frac{-\zeta_+^m}{\bar{w}(\zeta_+ - \zeta_-)} = \frac{\zeta_+^m}{2 r}
\end{equation}
so in particular we obtain the Coulomb field in the region $z>0$ when $m=0$. The reader is invited to explore the other cases. 

\section{The Vector Theory}
We now turn back to the treatment of the full vector potential theory. We need to elucidate the relationship between the vector potential approach and the more traditional use of stream functions. In the case of planar 2D flow, there is almost nothing remaining to do in the biharmonic case. If we set the vector potential to be
\begin{equation}
\Psivec = - \psi(x,y) \underline{e}_z
\end{equation}
the divergence-free condition is satisfied identically and we merely need that 
\begin{equation}
\biggl(\frac{\partial^2 \ }{\partial x^2} + \frac{\partial^2 \ }{\partial y^2} \biggr)^2 \psi = 0
\end{equation}
and this is all taken care of by the representation of \eqref{biharsoltwo}. Note that this well-known intrinsically 2D and superficially non-holomorphic result is now clearly understood in 3D {\it holomorphic} terms.
\subsection{Axis-symmetric vector flow}
This is traditionally modelled (Milne-Thomson, 1996) in terms of the Stokes stream function $\Psi_{S}(r, \theta)$. The components of the velocity field are given by
\begin{equation}
u_r = \frac{1}{r^2 \sin(\theta)}\frac{\partial \Psi_{S}}{\partial \theta}\ ,\ \ u_{\theta} = -\frac{1}{r \sin(\theta)}\frac{\partial \Psi_{S}}{\partial r}
\end{equation}
What this representation is {\it really} telling us, as is made clear in modern fluid theory, is that the {\it vector} potential $\Psivec$ for the flow is given by
\begin{equation}
\Psivec = \frac{\Psi_S}{r \sin(\theta)} \underline{e}_{\phi}
\end{equation}
as is revealed, together with the fact that $\Psivec$ is divergence-free, by elementary calculations with the curl and div operator expressed in a spherical basis. A further elementary calculation in vector calculus shows that for an axis-symmetric function $f(r, \theta)$
\begin{equation}
\del \times (\del \times \frac{f}{r\sin(\theta)}\underline{e}_{\phi}) = \frac{-1}{r\sin(\theta)}(E^2 f) \underline{e}_{\phi}
\end{equation}
where the operator $E^2$ is given by

\begin{equation}
E^2 f = \frac{\partial^2 f}{\partial r^2} + \frac{\sin(\theta)}{r^2}\frac{\partial \ }{\partial \theta}\biggl(\frac{1}{\sin(\theta)}\frac{\partial f}{\partial \theta} \biggr )
\end{equation}
The biharmonic condition may be expressed as the scalar PDE
\begin{equation}
E^4 \Psi_S = 0
\end{equation}
However, this representation in some ways obscures the underlying simplicity of the problem. To see why, we need to work with the problem in the full vector form, and, perhaps surprisingly, recast it in a Cartesian basis. In this way we can use the contour integral technology already developed for the scalar biharmonic problem. We write the basis vector $\underline{e}_{\phi}$ in the form
\begin{equation}
\underline{e}_{\phi} = \frac{-y \exvec + x \eyvec}{\sqrt{x^2 + y^2}} = \frac{-y \exvec + x \eyvec}{r \sin(\theta)} = \frac{1}{r \sin(\theta)} \Re\{w[i\exvec + \eyvec]\}
\end{equation}
where $w = x+i y$ as before. We could equally well write this down in terms of $\bar{w}$. Now recall that the full vector potential is given in terms of the Stokes stream function by the relation
\begin{equation}
\Psivec = \frac{\Psi_S}{r \sin(\theta)} \underline{e}_\phi
\end{equation}
and if $\Psi_S$ is real we can write the vector potential as
\begin{equation}
\Psivec = \Re\biggl\{ \frac{\Psi_S w}{r^2 \sin^2(\theta)} [i\exvec + \eyvec] \biggr \}
\end{equation}
The components of this with respect to a Cartesian basis must satisfy the scalar biharmonic equation, or indeed, as a special case, the Laplace equation. We now appeal to equation (3.36), where we note that $\psi_-$ is just a harmonic function. It follows that we can write the parts of $\Psi_S$ that are biharmonic but not harmonic in the form
\begin{equation}
\Psi_S = r^2 \sin^2(\theta) g(r, \theta) = w \bar{w} \frac{1}{2 \pi i} \int d \zeta  \frac{1}{\zeta} f\biggl(\frac{\eta}{\zeta}\biggr) 
\end{equation} 
for some complex function $f$. Here $g$ is harmonic and axis-symmetric and can therefore be written in terms of the $\psi_{n,0}$ functions given in equation (3.44) or indeed in terms of normal Legendre functions and powers of $r$ via a Laurent expansion of $f$ in the form 
\begin{equation}
\Psi_S = r^2 \sin^2(\theta) g(r, \theta) = w \bar{w} \frac{1}{2 \pi i} \int d \zeta  \sum_{n=-\infty}^{\infty} \frac{a_n}{\zeta^{n+1}}(w+2z\zeta-\bar{w} \zeta^2)^n
\end{equation} 
We argue that these relations are the natural axis-symmetric versions of (3.34). Of course, in general, we need to add in harmonic components, just as in the 2D planar case where we can add to $\Psi$ any pair $k_1(w)$ and $k_2(\bar{w})$ of holomorphic and anti-holomorphic functions. To treat this we look again at the representation in a Cartesian basis, this time in the form:
\begin{equation}
\Psivec = \frac{\Psi_S}{r \sin(\theta)} (-\sin(\phi) \exvec + \cos(\phi) \eyvec)
\end{equation}
We deduce that the function
\begin{equation}
\frac{\Psi_S}{r\sin(\theta)} e^{\pm i \phi}
\end{equation}
must be harmonic and therefore a solution of Laplace's equation with $m=\pm 1$ as described above. By packaging this up as before, we can write these harmonic contributions to $\Psi_S$, say $\Psi_{SH}$ in the elegant form
\begin{equation}
\Psi_{SH} = \frac{\bar{w}}{2\pi i}\int d\zeta \beta\biggl(\frac{\eta}{\zeta}\biggr) + \frac{w}{2\pi i}\int d\zeta \frac{1}{\zeta^2} \gamma\biggl(\frac{\eta}{\zeta}\biggr)
\end{equation}
for some choice of complex functions $\beta$ and $\gamma$. 

We are finally able to give the proposed contour integral solution for the Stokes stream function for axis-symmetric biharmonic flow, by combining these expressions into the form:
\begin{equation}
\Psi_S = w \bar{w} \frac{1}{2 \pi i} \int d \zeta  \frac{1}{\zeta} f\biggl(\frac{\eta}{\zeta}\biggr) +\frac{\bar{w}}{2\pi i}\int d\zeta \beta\biggl(\frac{\eta}{\zeta}\biggr) + \frac{w}{2\pi i}\int d\zeta \frac{1}{\zeta^2} \gamma\biggl(\frac{\eta}{\zeta}\biggr)
\end{equation} 
where $\eta$ is written in terms of $x,y,z$ and where $f, \beta, \gamma$ have Laurent series expansions that generate expansions in terms of powers of $r$ and regular ($f$) and modified ($\beta,\gamma$) functions. 

\subsection{A simple example to check it all works}
We can see that a choice of $f$ constant gives a contribution to $\Psi_S$ proportional to 
\begin{equation}
w\bar{w} = r^2 \sin^2(\theta)
\end{equation}
We also know (at least locally) that the choice $f(z) = 1/z$ gives a Coulomb field and a contribution to $\Psi_S$ proportional to 
\begin{equation}
\frac{w\bar{w}}{r} = r \sin^2(\theta) 
\end{equation}
Another interesting contribution can be generated by the choice $\beta(z) = 1/z^2$, where an elementary exercise in the calculus of residues leads to a contribution to $\Psi_S$ of the form
\begin{equation}
\frac{w\bar{w}}{r^3} = \frac{1}{r} \sin^2(\theta) 
\end{equation}
If we take a general linear combination of these three in the form
\begin{equation}
\sin^2(\theta)\biggl\{A r^2 + B r + \frac{C}{r} \biggr\}
\end{equation}
we obtain a valid stream function. The particular choice
\begin{equation}
\Psi_S = \frac{U}{2}\sin^2(\theta)\biggl\{r^2 -\frac{3 a r}{2} + \frac{a^3}{2 r} \biggr\}
\end{equation}
gives the well-known stream function for very viscous flow around a sphere of radius $a$ and uniform flow at at rate $U$ at infinity. Having non-dimensionalized we would, for example, scale so that $a=1$. In general we have a contour integral technique for solving the PDE given by equation (4.7). We should note of course that this discussion is confined to showing only that the contour integrals contain the well known historical solution for a sphere due to Stokes. The problems associated with this solution are also well known - see for example Chapter 8 van Dyke (1964) for a discussion of the {\it Whitehead paradox} and its resolution by matched asymptotic expansions.

\section{Small but non-vanishing Reynolds number}
A natural question to ask is to wonder how much of the above is dependent on the purely linear structure that arises in the biharmonic limit? We cannot yet answer this question for a general Reynolds number in three dimensions, but we can observe that something very interesting happens when we (a) go back to two dimensions and (b) consider the case of a small but non-zero Reynolds number. Let us go back to the non-dimensional form of Eq. (2.13). This is 
\begin{equation}
\nabla^4 \psi = R \frac{\partial(\psi, \delsq \psi)}{\partial(x, y)}
\end{equation}
In terms of the complex variable $w = x + i y$, we can write this in the form
\begin{equation}
i \frac{\partial^4 \psi}{\partial w^2 \partial \bar{w}^2} = \frac{R}{2} \biggl(
\frac{\partial \psi}{\partial w}\frac{\partial^3 \psi}{\partial w \partial \bar{w}^2} - 
\frac{\partial \psi}{\partial \bar{w}}\frac{\partial^3 \psi}{\partial \bar{w} \partial w^2}
\biggr)
\end{equation} 
Rather than pursuing the approach of Legendre (1949) and Ranger (1991, 1994) we can consider instead the interesting physical case of small but non-vanishing Reynolds number. Let us {\it assume} that the solution for $\psi$ may be written as
\begin{equation}
\psi = \psi_0 + R \psi_1 + O(R^2)
\end{equation}
and that
\begin{equation}
\psi_0 = \Re\bigl\{\bar{w} f_0(w) + g_0(w)\bigr\}
\end{equation}
This is a very strong assumption, and it is well known that this assumption of a power series dependence on the Reynolds number may fail. There may not indeed be a sensible form for $\psi_0$ over a simple domain of interest. The reader is again referred to Chapter 8 of van Dyke (1964) for a discussion of the {\it Stokes paradox} for a cylinder in a uniform flow. Our purpose here is to illustrate that the low Reynolds number perturbation equation may indeed be integrated using holomorphic methods. The result may be of use in refining the results for a certain sub-class of problems where there is both a meaningful $\psi_0$ {\it and} the inertia terms in the Navier-Stokes equations (i.e. the non-linear terms) arising from $\psi_0$ remain small over the entire domain of interest. Under these strong assumptions we can proceed.
The equation for $\psi_1$ is, under these assumptions,
\begin{equation}
i \frac{\partial^4 \psi_1}{\partial w^2 \partial \bar{w}^2} = \frac{1}{8} \biggl(
\overline{f_0^{''}(w)}\bigl[ \bar{w}f'_0(w) + g'_0(w) + \overline{f_0(w)} \bigr] 
- f_0^{''}(w)\bigl[ w \overline{f'_0(w)} + \overline{g'_0(w)} + f_0(w) \bigr] 
\biggr)
\end{equation} 
This may be solved almost explicitly as follows. We let $F(w),G(w),H(w)$ be {\it holomorphic} functions with the properties
\begin{equation}
F'(w) = f_0(w)\ ,\ \ \    G'(w) = g_0(w)\ ,\ \ \ H''(w) = f_0(w)f''_0(w) 
\end{equation}
Then a particular solution to Eq. (5.5) is given by
\begin{equation}
\psi_{1P} = \frac{1}{4} \biggl(
\overline{(w F'(w) - 2 F(w))}F(w) + \overline{F'(w)}G(w) + \frac{w^2}{2}\overline{H(w)}
\biggr)
\end{equation}
and a complementary function exists in the obvious form
\begin{equation}
\psi_{1CF} = \Re\bigl\{\bar{w} f_1(w) + g_1(w)\bigr\}
\end{equation}
where $f_1$ and $g_1$ are arbitrary holomorphic functions.
So we can see that apart from the practical issue on constructing the integrals in Eq. (5.6), the first perturbation can be constructed by separate integration of the $w$ and $\bar{w}$ components. In fact, we have shown that the perturbative non-linear problem may be solved in terms of free holomorphic functions $F,G,f_1,g_1$ and the solution, apart from the construction of $H$, is given explicitly in terms of this holomorphic information. This observation gives some hope that a corresponding three-dimensional structure might exist. 
\section{Summary}

We have presented the theory of 3D twistors in such a way as to allow the understanding of biharmonic flow in three dimensions. The representations developed are a natural generalization of the stream function method that is very familiar to fluid dynamicists.  Hopefully this work will stimulate the use of twistor methods in dealing with 3D problems in applied mathematics and fluid dynamics in particular. The ultimate goal is a better understanding of the Navier-Stokes equations through the geometry of complex variable techniques. Ideally we would be able to draw together the threads of work given by Legendre (1949),  Ranger (1991,1994), and more recently, Roulstone {\it et al} (2005). Another major issue is the imposition of boundary conditions directly in twistor space. But it is quite clear that we can generate elegant contour integral solutions to problems in fluid dynamics, beyond those already well known for two dimensions and potential flow. We have also seen that the case of small but non-zero Reynolds number is open to complex variable treatment in the case of two dimensions, though at present we need to make some strong assumptions on the nature of the flow. Finally the generality of the proposed representations needs to be established. So far we have shown how to construct an interesting class of solutions, but further work is needed on just `how large' this class is. 

\subsection*{Acknowledgements}
The author is grateful to  Dr. S.D Howison, Dr. L.J. Mason, Dr. I. Roulstone, Dr. J.R. Ockendon, F.R.S for conversations clarifying several issues related to this work.

\label{lastpage}
\end{document}